\documentclass[preprint,showpacs,preprintnumbers,amsmath,amssymb]{revtex4}

\usepackage{graphicx}
\usepackage{dcolumn}
\usepackage{bm}

\begin{document}


\title{Periodic-orbit determination of dynamical correlations \\ in stochastic processes}
Accepted for publication in Phys. Rev. E.
\author{Miki U.~Kobayashi}
 \email{miki@acs.i.kyoto-u.ac.jp}
\author{Hirokazu Fujisaka}
 \email{fujisaka@i.kyoto-u.ac.jp}
\author{Syuji Miyazaki}%
 \email{syuji@i.kyoto-u.ac.jp}
\affiliation{%
Department of Applied Analysis and Complex Dynamical Systems,\\
Graduate School of Informatics, Kyoto University,
Kyoto 606-8501, Japan
}%


\date{\today}

\begin{abstract}
It is shown that large deviation statistical 
quantities of the discrete time, finite state Markov process 
$P_{n+1}^{(j)}=\sum_{k=1}^NH_{jk}P_n^{(k)}$, where $P_n^{(j)}$ is the 
probability for the $j$-state at the time 
step $n$ and $H_{jk}$ is the transition probability,  
completely coincides with those from the 
Kalman map corresponding to the above Markov process.  
Furthermore, it is demonstrated that by using simple examples, time 
correlation functions in finite state Markov processes can be 
well described in terms of unstable periodic orbits embedded in 
the equivalent Kalman maps.
\end{abstract}

\maketitle

\section{Introduction}
Over the last three decades, nonlinear dynamics and chaos 
have played significant roles not only in natural science 
and mathematics but also in engineering and social 
science~\cite{BPV84,U92,SJ05,O93,MK97}.  
Chaotic dynamics has many different kinds of aspects.  
This is one of the reasons why many researchers in 
various fields have studied extensively for so long time.  
\par
The most important characteristic of chaos is 
the ``trajectory instability'', which makes the long-term 
predictability impossible~\cite{BPV84,SJ05,O93,L63}.  This 
characteristic can be quantified with positive Lyapunov exponent.  
Owing to the long-term unpredictability, chaotic 
dynamics can be used as random number generators.  
On the other hand, one cannot predict the state of a physical 
variable in future obeying a stochastic process.  However, the origin 
of the unpredictability in stochastic process is believed to be different 
from that for chaotic dynamics.  In connection with this fact, 
one may ask ``Is it possible to precisely simulate stochastic process 
with a chaotic dynamics suitably constructed?''.  In other words, 
does a hidden dynamics which gives the same statistics as the 
stochastic process under consideration exist?  The 
present paper concerns with this question.
\par
The possibility of the construction of the one-dimensional chaotic 
map which is equivalent to a finite-state Markov process
\begin{eqnarray}
P_{n+1}^{(j)}=\sum_{k=1}^NH_{jk}P_n^{(k)},
\end{eqnarray}
$(n=0,1,2,3,\cdots )$  was proposed by Kalman~\cite{K57} in 1956.  
Here, $P_n^{(j)}$ is the probability that 
the state is in the $j$-th state at 
time step $n$, ($\sum_{j=1}^NP_n^{(j)}=1$).  
The $H_{jk}$ is the transition probability 
from the $k$-th state to the $j$-th state in a time step 
and satisfies $\sum_{j=1}^NH_{jk}=1$.  
Kalman showed how to construct one-dimensional chaotic map
\begin{eqnarray}
x_{n+1}=f(x_n)
\end{eqnarray}
corresponding to the process (1), 
where the mapping function $f(x)$ is completely determined by the 
transition matrix $\hat{H}=\{ H_{jk} \} $.  The mapping dynamics 
(2) constructed by Eq.~(1) is called the {\it Kalman map} 
and its simplified review is given in Appendix A.  
The Kalman map is a sort of piecewise linear Markov transformations 
which map each interval of the partition onto a union of intervals of the partition.
Markov transformations have been studied by many researchers~\cite{R1957,BG1997,D1999} and play an important role in the study of chaotic dynamics.
In fact, Kalman map gives the invariant probability 
same as the the corresponding 
Markov map (1).  Recently, Kohda and Fujisaki~\cite{KF99} showed that 
the double-time correlation function  
obtained from (2) precisely agrees with that 
obtained from the corresponding Kalman map.
\par  
In the sense that the stochastic process is generated 
by the deterministic chaotic dynamics, the latter may be called the 
{\it hidden dynamics of stochastic process}.  This fact leads to a quite 
interesting problem.  It is well known that statistical quantities including 
dynamical correlation functions in chaotic dynamics can be well 
approximated in terms of unstable periodic orbits embedded in the corresponding 
strange attractor.  Therefore, if a stochastic process can be determined by 
a corresponding chaotic dynamics, it is naturally expected that statistical 
quantities in a stochastic process can be determined by the corresponding 
chaotic dynamics, particularly in terms of unstable periodic orbits embedded 
in the latter dynamics and, therefore, in the stochastic dynamics.  The main aim of the 
present paper is to show the possibility of the determination of the 
statistics of the stochastic process (1) in terms of unstable periodic orbits 
of the corresponding Kalman map.  Namely, the present paper concerns with the 
hidden dynamics of a finite-state, discrete-time stochastic 
process and, in particular, the relation between statistical properties of 
stochastic process 
and unstable periodic orbits being subtended in the hidden dynamics.
The main results of the present paper are as follows:
\begin{enumerate}
\item[(i)] The large deviation statistical quantities   
calculated in the stochastic process (1) rigorously coincide with 
those derived from the corresponding Kalman map.  
Furthermore, 
\item[(ii)] by using several simple stochastic processes, we show 
that double-time correlation functions of stochastic processes can be well 
approximated with those obtained in terms of unstable periodic orbits embedded 
in the corresponding Kalman map.
\end{enumerate}  
\par
The present paper is organized as follows.  
In Sec.~II, we briefly summarize statistical quantities of the process (1) 
and the construction of the corresponding Kalman map.  It is shown that the 
statistical quantities, i.e., the invariant density, the double-time 
correlation function and the large deviation statistical quantity for the 
stochastic process rigorously coincide with those obtained from the Kalman 
map.  In Sec.~III, making use of simple models, we will show that the 
double-time correlation function of the stochastic process can be well 
approximated by those determined in terms of unstable periodic orbits 
embedded in the Kalman map.  Concluding remarks and discussions are given 
in Sec.~IV.  In Appendix A, explicit mapping functions are given for two- 
and three-states stochastic processes.  Appendix B is devoted to the proof 
of the statements in Sec.~II.  In Appendix C, we will briefly review the 
Markov method which enables the double-time correlation function to be 
expanded in terms of same-time correlation functions.  Furthermore, in 
Appendix D, we give a comparison of the double-time correlation function 
determined with a single unstable periodic orbit.

\section{Equivalence of statistical dynamics generated by stochastic 
process and the Kalman map}
In the matrix form, the Markov process (1)
is written as
\begin{eqnarray}
{\bm P}_{n+1}=\hat{H}{\bm P}_n,
\end{eqnarray}
where
$
{\mbox{\boldmath{$P$}}}_n=(P_n^{(1)}, P_n^{(2)},\cdots ,P_n^{(N)})^T
$
is the probability matrix and $\hat{H}$ is the transition matrix
with the $jk$ element $H_{jk}$. 
The steady probability distribution is determined by
\begin{eqnarray}
{\bm P}_{*}=\hat{H}{\bm P}_{*},
\end{eqnarray}
where
$
{\bm P}_{*}=(P_*^{(1)}, P_*^{(2)},\cdots ,P_*^{(N)})^T.
$ 
This implies that ${\mbox{\boldmath{$P$}}}_{*}$ is the eigenvector of
$\hat{H}$ with the eigenvalue 1.
\par
Let us consider the dynamical variable $u_n$ at the time step $n$, and
let $u_n$ take the value $a^{(j)}$ if the state is in the $j$-th state. 
The double-time correlation function for the fluctuation $\delta
u_n=u_n-
\langle u\rangle $, $\langle u \rangle (=\sum_{j=1}^N
a^{(j)}P_*^{(j)})$ being the average value, is given by
\begin{eqnarray}
C_n&=&\langle \delta u_n\delta u_0\rangle  \nonumber \\
&=& \sum_{j=1}^N\sum_{k=1}^N\delta a^{(k)}(\hat{H}^n)_{kj}\delta
a^{(j)}
P_*^{(j)}
\end{eqnarray}
with $\delta a^{(j)}=a^{(j)}-\langle u\rangle $.
\par
Furthermore, for the sake of later discussion, we here briefly summarize
the large deviation theoretical study of the stochastic
process.  Consider a steady time series $\{ u_j\}$.  The finite-time
average
\begin{eqnarray}
\overline{u}_n=\frac{1}{n}\sum_{j=0}^{n-1}u_j,
\end{eqnarray}
$n$ being the time span of averaging, is a fluctuating variable. 
For $n\rightarrow \infty $, the average approaches the ensemble
average $\langle u\rangle $.  However, for a large but finite $n$,
$\overline{u}_n$
shows a fluctuation.  Let $Q_n(u)$ be the probability density that
$\overline{u}_n$ takes the value $u$.  As is known, for a large $n$,
$Q_n(u)$ asymptotically takes the form $Q_n(u)\sim
e^{-S(u)n}$~\cite{E85},
where $S(u)$ called the {\it rate function} or the {\it fluctuation
spectrum}~\cite{FI87a}
is a concave function of
$u$ and has a minimum $S=0$ at $u=\langle u \rangle $. 
The fluctuation spectrum characterizes the fluctuation statistics
of the time series $\{ u_k\} $. 
The large deviation theoretical characteristic function
$Z_q(n)$ for the time series $\{ u_k\} $ is defined by
\begin{eqnarray}
Z_q(n)=\langle \exp (qn\overline{u}_n)\rangle
= \Bigg< \exp \left( q\sum_{k=0}^{n-1}u_k \right)  \Bigg> ,
\end{eqnarray}
where $q$ is an arbitrary real number and $\langle \cdots \rangle $ is
the
ensemble average.  For a large $n$, $Z_q(n)$ asymptotically takes the
form $Z_q(n)\sim e^{\phi (q)n}$, where $\phi (q)$ depends only on the
parameter $q$ and characterizes the statistics of $\{ u_k \} $. 
The fluctuation spectrum $S(u)$ is derived by the Legendre transform of
$\phi (q)$ as $\phi
(q)=-\min_{u}[S(u)-qu]$~\cite{FI87a,FI87b,FI89,FS91,F92,F05,JF93}.
\par
If we introduce the generalized transition matrix $\hat{H}_q$ with its
$jk$ element defined via~\cite{FI89}
\begin{eqnarray}
\left( \hat{H}_q\right)_{jk}=H_{jk}\exp \left[ qa^{(k)} \right] .
\end{eqnarray}
The large deviation theoretical characteristic function is written as
\begin{eqnarray}
Z_q(n)=\sum_{k=1}^N\left( \hat{H}_q^n
{\mbox{\boldmath{$P$}}}_{*}\right)_k.
\end{eqnarray}
Therefore, one finds that the characteristic function $\phi (q)$ is
determined by the largest eigenvalue of the generalized transition
matrix
$\hat{H}_q$~\cite{FI89}.
\par
It is known that the same steady probability distribution as that
in Eq.~(4) can be
produced by a one-dimensional map, called the Kalman map, suitably
constructed.  In the remaining part of this section, we first
explain how to construct the Kalman map.  The following discussion is
the
simplified one proposed by Kalman~\cite{K57}.
\par 
First, we define the positions $\beta_{\underline{j}}\
(\underline{j}=1,2,\cdots ,N^2)$
with $\beta_0=0$ and $\beta_{N^2}=1$ in such a way that they satisfy
\begin{eqnarray}
\frac{\beta_N}{\beta_{(j-1)N+1}-\beta_{(j-1)N}}&=&H_{1j}^{-1}
,\nonumber \\
\frac{\beta_{2N}-\beta_N}{\beta_{(j-1)N+2}-\beta_{(j-1)N+1}}&=&H_{2j}^{
-1} ,\nonumber \\
\frac{\beta_{3N}-\beta_{2N}}{\beta_{(j-1)N+3}-\beta_{(j-1)N+2}}&=&H_{3j
}^{-1} , \nonumber \\
&\vdots &  \\
\frac{\beta_{kN}-\beta_{(k-1)N}}{\beta_{(j-1)N+k}-\beta_{(j-1)N+k-1}}&=
&H_{kj}^{-1} ,\nonumber \\
&\vdots & \nonumber \\
\frac{\beta_{N\cdot
N}-\beta_{(N-1)N}}{\beta_{(j-1)N+N}-\beta_{(j-1)N+N-1}}&=&H_{Nj}^{-1},
\nonumber
\end{eqnarray}
$(j=1,2,\cdots ,N)$. 
These equations are solved to yield
\begin{eqnarray}
\beta_{(j-1)N+k}=\frac{1}{N}(j-1+\sum_{\ell =1}^kH_{\ell j}),
\end{eqnarray}
$(k=0,1,2,\cdots , N)$.  We thus find that
$\beta_{(j-1)N+k}-\beta_{(j-1)N+k-1}=H_{kj}/N$. 
With these $\beta_{\underline {j}}$, we construct the piecewise linear
one-dimensional map $f(x)$
in the range $\beta_{(j-1)N}<x\le \beta_{jN}$, $(j=1,2,\cdots ,N)$ as
follows:
\begin{eqnarray}
f(x)=
      \begin{cases}
      H_{1j}^{-1}(x-\beta_{(j-1)N}) & \text{($\beta_{(j-1)N}< x\le
\beta_{(j-1)N+1}$)}, \\
      \beta_{N}+H_{2j}^{-1}(x-\beta_{(j-1)N+1}) &
\text{($\beta_{(j-1)N+1}< x\le \beta_{(j-1)N+2}$)}, \\
      \beta_{2N}+H_{3j}^{-1}(x-\beta_{(j-1)N+2}) &
\text{($\beta_{(j-1)N+2}< x\le \beta_{(j-1)N+3}$)} ,\\
       \hspace{3cm} \vdots & \text{\ \ } \\     
      \beta_{(k-1)N}+H_{kj}^{-1}(x-\beta_{(j-1)N+k-1}) &
\text{($\beta_{(j-1)N+k-1}< x\le \beta_{(j-1)N+k}$)}, \\
       \hspace{3cm} \vdots & \text{\ \ } \\ 
      \beta_{(N-1)N}+H_{Nj}^{-1}(x-\beta_{(j-1)N+N-1}) &
\text{($\beta_{(j-1)N+N-1}< x\le \beta_{(j-1)N+N}$)}.
      \end{cases} \nonumber \\
\end{eqnarray}
Therefore, one finds $|f'(x)|^{-1}=H_{kj}$ for $\beta_{(j-1)N+k-1}<
x\le \beta_{(j-1)N+k}$. 
Examples for $N=2$ and 3 are shown in Appendix A. 
One should note that the dynamics $x_{n+1}=f(x_n)$ with the mapping
function (12)
shows a chaotic behavior since the local expansion rate $\ln |f'(x)|$ of
the
mapping function is everywhere positive.  Therefore, the mapping
system (12) turns out to be hyperbolic.
\par
Let $u(x_n)$ be a dynamical variable taking the value $u(x)=a^{(j)}$
if
$x$ satisfies $\beta_{(j-1)N}<x\le \beta_{jN},\ (j=1,2,\cdots ,N)$. 
The time correlation function $C_n^{\rm K}$ of $u(x_n)$ is
given by
\begin{eqnarray}
C_n^{\rm K}= \langle \delta u(x_n)\delta u(x)\rangle ,
\end{eqnarray}
where $\langle \cdots \rangle $ is the ensemble average. 
Here, $x_n=f^n(x)$ and $\delta u(x_n)=u(x_n)-\langle u\rangle $. 
The large deviation statistical characteristic function is
defined by
\begin{eqnarray}
Z_q^{\rm K}(n)
= \Big< \exp \left[ q\sum_{m=0}^{n-1}u(x_m) \right]  \Big> .
\end{eqnarray}
As is shown in Appendix B, one finds that $C_n=C_n^{\rm K}$
and $Z_q(n)=Z_q^{\rm K}(n)$.
\par
It should be noted that dynamical statistical quantities are determined
in
terms of unstable periodic orbits embedded in chaotic dynamics~\cite{KF06}.  
This facts implies that since the Markov stochastic process (1) is 
described by the corresponding Kalman chaotic dynamics, dynamical statistical
quantities
such as time correlation functions and large deviation theoretical
statistical
characteristic functions of the Markov stochastic process (1) can be
determined by unstable periodic orbits embedded in the ``stochastic
process''. 
In the following section, we discuss the determination of dynamical
quantities of
the stochastic process in terms of periodic orbits embedded in the
Kalman map.
\section{Periodic-orbit determination of time correlation functions
in simple stochastic processes}
In this section, we will show that the time correlation functions $C_n$'s 
for stochastic processes with $N=2$ and $3$ are well determined by 
unstable periodic orbits embedded in the chaotic dynamics 
corresponding to the stochastic processes.  
From Eq.~(5), the time correlation function for the 
stochastic process can be rigorously obtained as follows.  
Let $\mu_{\alpha }$ and ${\mbox{\boldmath{$e$}}}_{\alpha }$ be 
respectively the $\alpha $-th eigenvalue and the eigenvector 
of $\hat{H}$, i.e., $\hat{H}{\mbox{\boldmath{$e$}}}_{\alpha }=
\mu_{\alpha }{\mbox{\boldmath{$e$}}}_{\alpha }$.  
One would note that there is one eigenstate with the eigenvalue $\mu_{\alpha}=1$.
Without losing generality, we put $\mu_1 = 1$.
With the 
expansion $\delta a^{(j)}P_*^{(j)}=\sum_{\alpha }'b_{\alpha }
e_{\alpha }^{(j)}$, $e_{\alpha }^{(j)}$ being the $j$-th 
component of ${\mbox{\boldmath{$e$}}}_{\alpha }$, with the 
expansion coefficient $b_{\alpha }$, we obtain
\begin{eqnarray}
C_n={\sum_{\alpha }}'g_{\alpha }\mu_{\alpha }^n
\label{eq:exact-correlation}
\end{eqnarray}
with $g_{\alpha }=b_{\alpha }\sum_{j=1}^N\delta a^{(j)}
e_{\alpha }^{(j)}$, where $\sum_{\alpha }'$ implies the summation 
except the eigenvalue $\mu_1 =1$.
\par
As shown in Appendix C, the time correlation function can be approximately 
obtained as follows.  First, introduce the vector variable
\begin{eqnarray}
{\mbox{\boldmath{$u$}}}_n\equiv (h_1(x_n),h_2(x_n), 
\cdots ,h_{M+1}(x_n))^T\equiv {\mbox{\boldmath{$u$}}}\{ x_n\} ,
\end{eqnarray}
where $u_n\equiv h_1(x_n)=\sum_{j=1}^{N}a^{(j)}I_j(x_n)$ with 
the step function $I_j(x)$ defined as
\begin{eqnarray}
\underline{I}_{\underline{j}}(x)=
\begin{cases}
1 &  \text{for $\beta_{\underline{j}-1}<x\le \beta_{\underline{j} }$},\\
0 &  \text{otherwise}.
\end{cases}
\end{eqnarray}
And $h_j(x) =h_1(f^{j-1}(x))$, $(j=1,2,\cdots ,M+1)$.  
The time correlation function $C_n$ is given by the 1-1 
element $(\hat{C}_n)_{1,1}$ of the correlation matrix $\hat{C}_n\equiv 
\langle {\mbox{\boldmath{$u$}}}_n{\mbox{\boldmath{$u$}}}_0^T\rangle 
 -\langle {\mbox{\boldmath{$u$}}}_0\rangle 
\langle {\mbox{\boldmath{$u$}}}_0^T\rangle $.  If $M$ is suitably chosen, 
the matrix $\hat{C}_n$ is approximately determined by  
$\hat{C}_n\approx \hat{\zeta }^n\hat{C}_0$, where by noting $\hat{\zeta }=\hat{C}_1\hat{C}_0^{-1}$~\cite{F05,KF06}.   
Thus the time correlation 
function $C_n$ is determined by $\hat{C}_0$ and $\hat{C}_1$.
This method is referred to as the {\it Markov method} (Appendix C).
It should be noted that for a piecewise constant function 
$G(x)=\sum_{j=1}^NG_jI_j(x)$, its long time average is replaced by the 
ensemble average: 
\begin{eqnarray}
\langle G(x)\rangle \equiv \lim_{n\rightarrow \infty }
\frac{1}{n}\sum_{j=1}^nG(f^{j-1}(x))=\sum_{k=1}^NG_kP_*^{(k)}
\label{eq:average}
\end{eqnarray}
From definition, the functions ${\mbox{\boldmath{$u$}}}\{ x\}$ 
and ${\mbox{\boldmath{$u$}}}\{ f(x)\}$ which are relevant to the calculation of $\hat{C}_0$ and $\hat{C}_1$ 
are piecewise constant in the $x$ space.  Therefore $\hat{C}_0$ and 
$\hat{C}_1$ are obtained as the ensemble averages as in Eq.~(\ref{eq:average}).
\par
On the other hand, 
we can determine the quantities $\hat{C}_0$ and $\hat{C}_1$ 
in terms of infinitely many unstable periodic orbits
that can describe the invariant density $\rho (x)$ \cite{GOY88,MHMHT88,CE91,CAD03}.
In particular, the invariant density of one-dimensional chaotic systems
can be obtained as follows \cite{KT80}
\begin{eqnarray}
\rho(x)=\lim_{n \rightarrow \infty} \sum_{j=1}^{N(n)}\omega_j^{(n)}\delta (x-p_j^{(n)})
\label{eq:upo-density}
\end{eqnarray}
with
$$
\omega_j^{(n)}={C_n \over |(f^{n})'(p_j^{(n)})|}, \ \ (j=1,2,\cdots,N(n)), 
$$
where $p_j^{(n)}$ is a fixed point satisfying $f^{n}(p_j^{(n)})=p_j^{(n)}$,
and $N(n)$ is the total number of the fixed points of $f^{n}$ 
and $C_n$ is the normalization constant.
Thus the dynamical correlation functions can be 
approximately expanded 
in terms of unstable periodic orbits.  
\par
\begin{figure*}[th]
\begin{center}
\begin{tabular}{cccc}
   \resizebox{60mm}{60mm}{\includegraphics{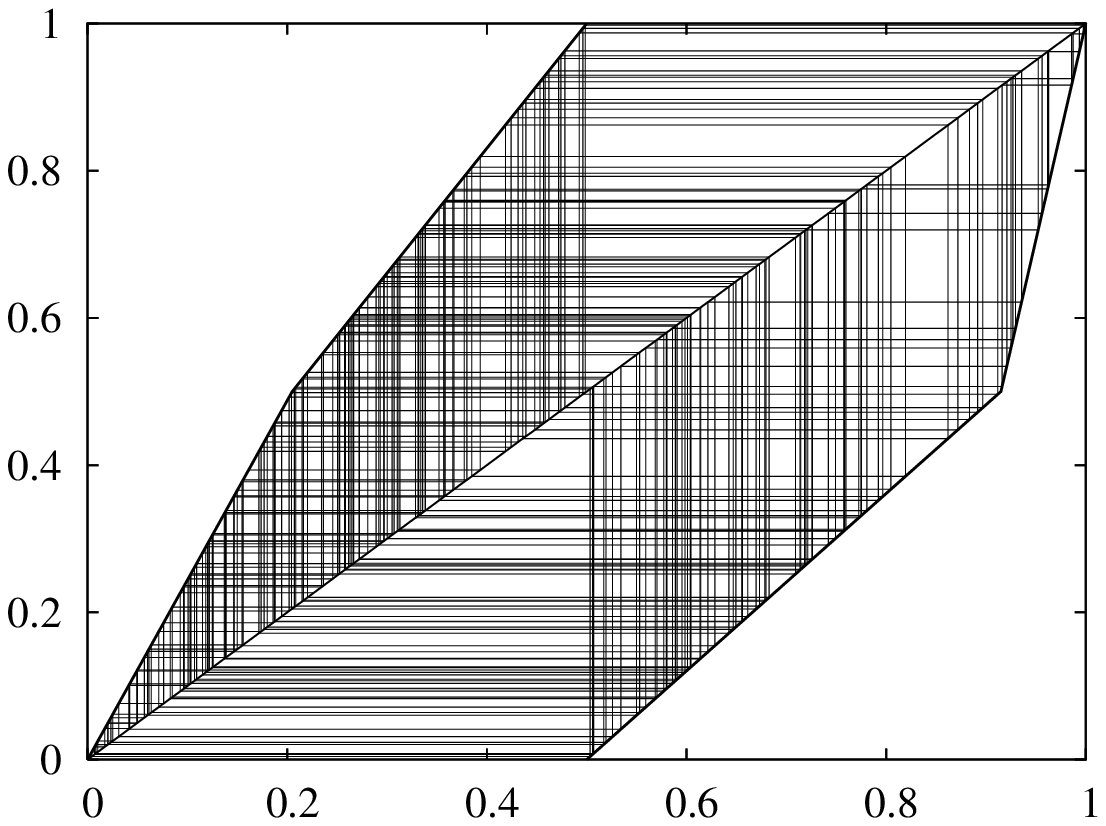}}&
   \resizebox{60mm}{60mm}{\includegraphics{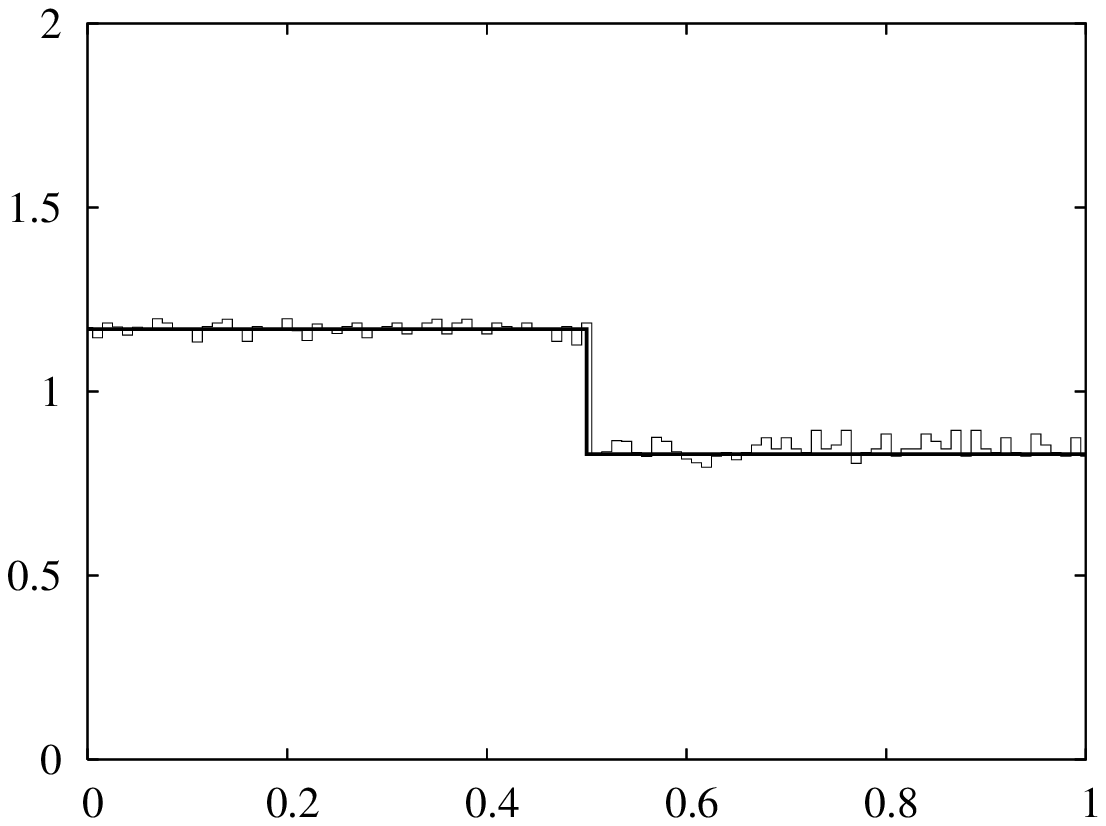}}\\
   \resizebox{60mm}{40mm}{\includegraphics{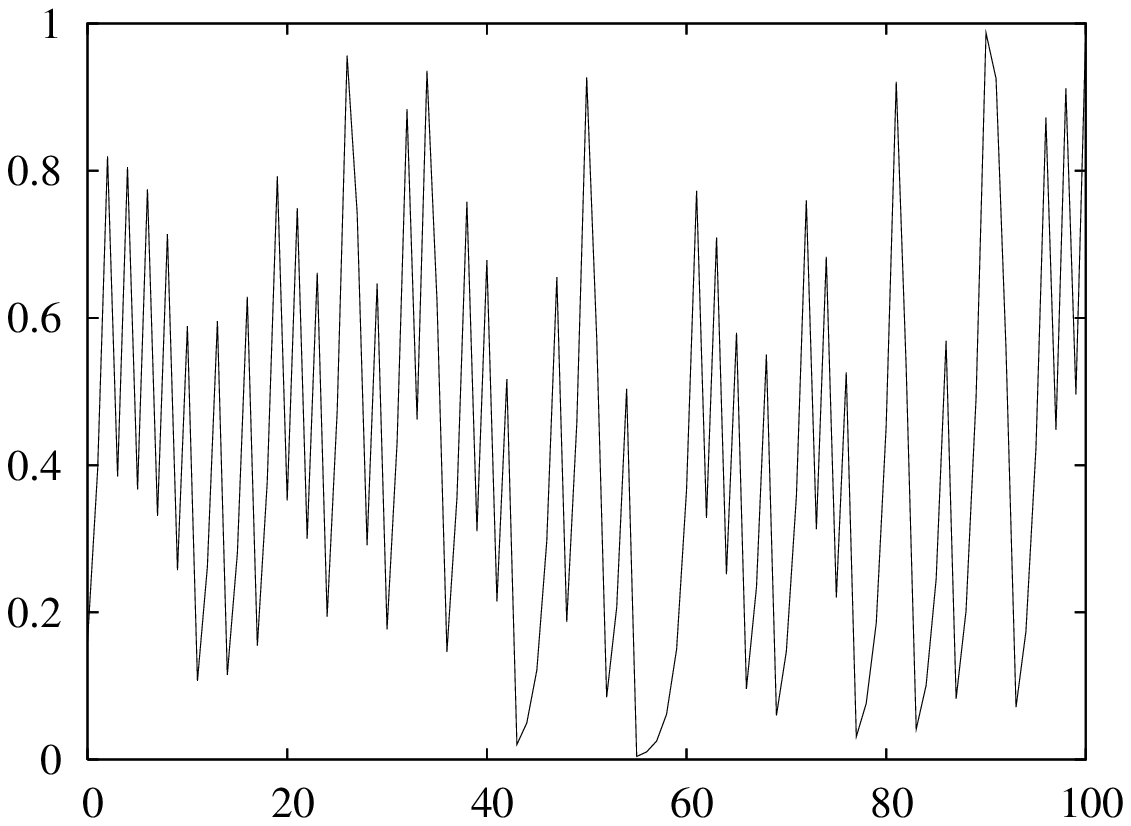}}&
   \resizebox{60mm}{40mm}{\includegraphics{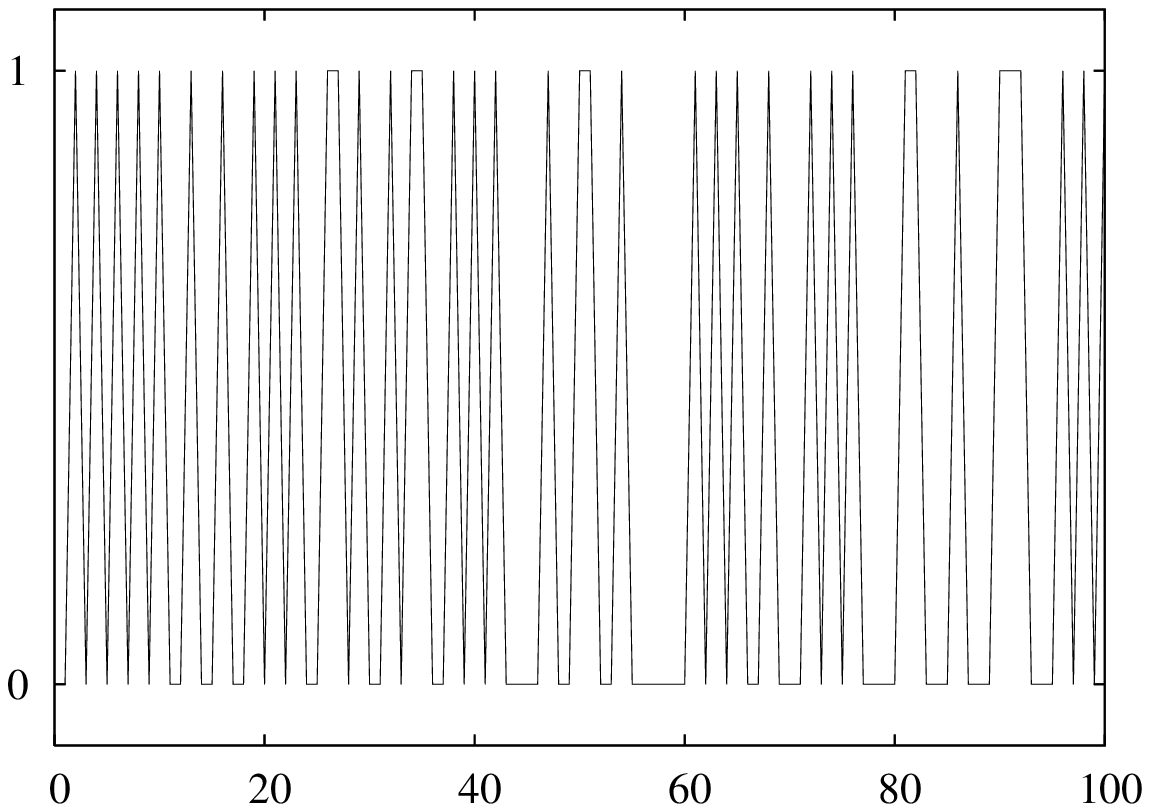}}\\
    \begin{picture}(0,0)(0,0)
      \put(0,310){(a)}
      \put(174,310){(b)}
      \put(-88,133){(c)}
      \put(5,18){{\large $n$}}
      \put(176,18){{\large $n$}}
      \put(88,224){\rotatebox{90}{\large $\rho(x)$}}
      \put(178,138){\large $x$}
      \put(5,138){\large $x_n$}
      \put(-88,70){\rotatebox{90}{\large $x_n$}}
      \put(90,70){\rotatebox{90}{\large $u_n$}}
      \put(-88,214){\rotatebox{90}{\large $x_{n+1}$}}
    \end{picture}
\end{tabular}
  \end{center}
  \caption{\label{fig:fig2}
Comparison of the  results by the Markov process and the Kalman map for $N=2$.  
(a) Kalman map corresponding to the Markov process with 
the transition matrix given in Eq.~(\ref{eq:H}).  
(b) Comparison of the exact invariant density (thick line)
with the approximate invariant density (thin line) in terms of unstable 15-periodic orbits by making use of the formula (\ref{eq:upo-density}).
(c) Chaotic time series generated by the mapping dynamics (Left) in (a) and 
the state evolution (Right) corresponding to the figure (a).  The stochastic 
evolution is generated by the equivalent Kalman dynamics.   
}
\end{figure*}
\begin{figure*}[th]
\begin{center}
\begin{tabular}{cccc}
   \resizebox{60mm}{60mm}{\includegraphics{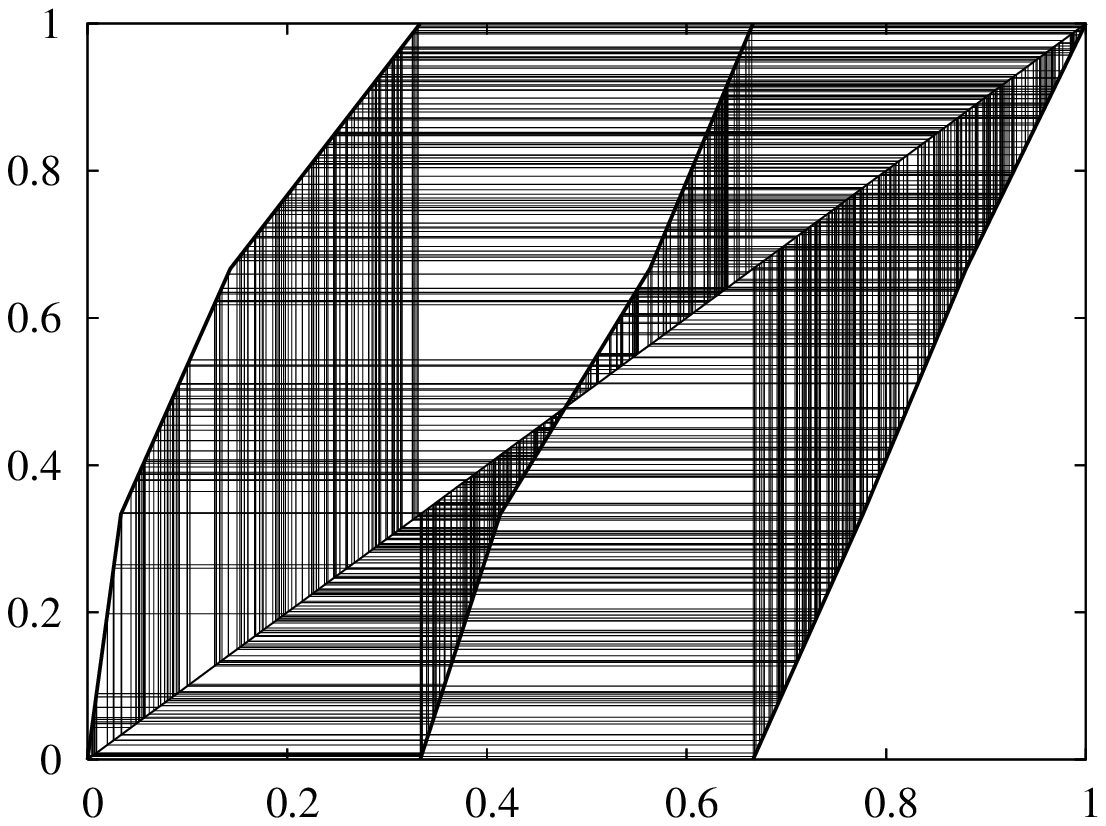}}&
   \resizebox{60mm}{60mm}{\includegraphics{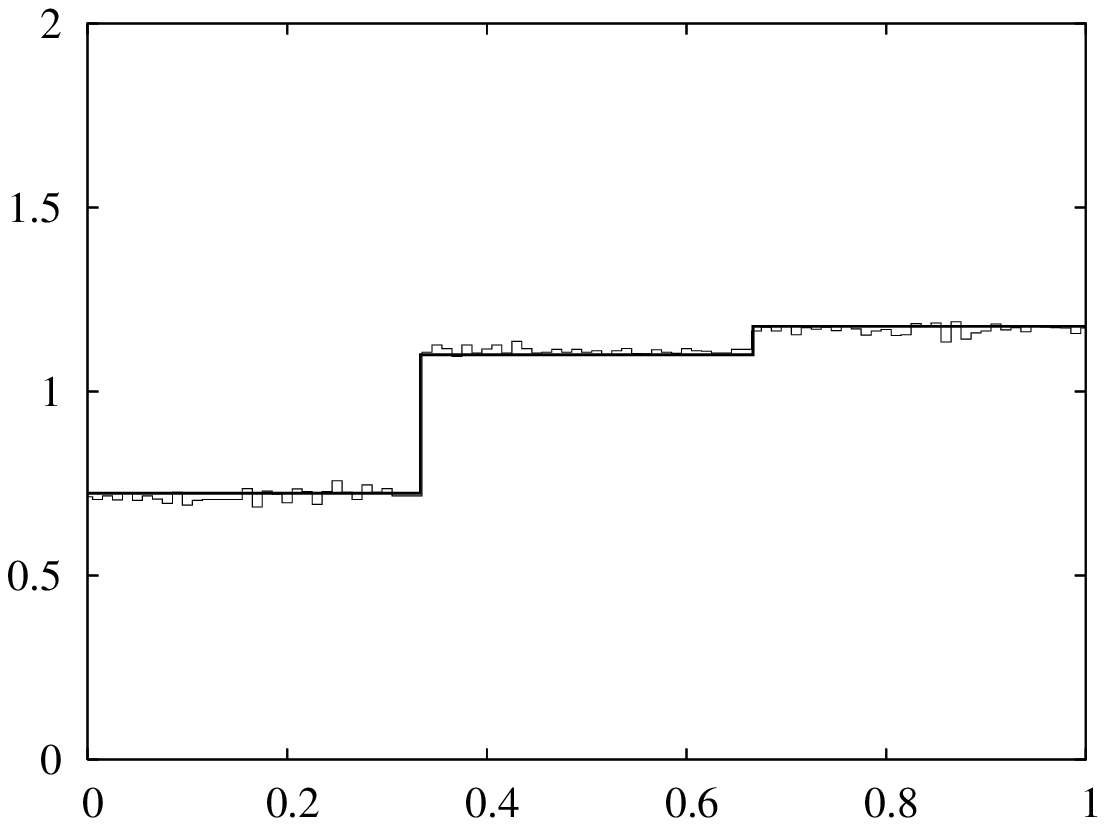}}\\
   \resizebox{60mm}{40mm}{\includegraphics{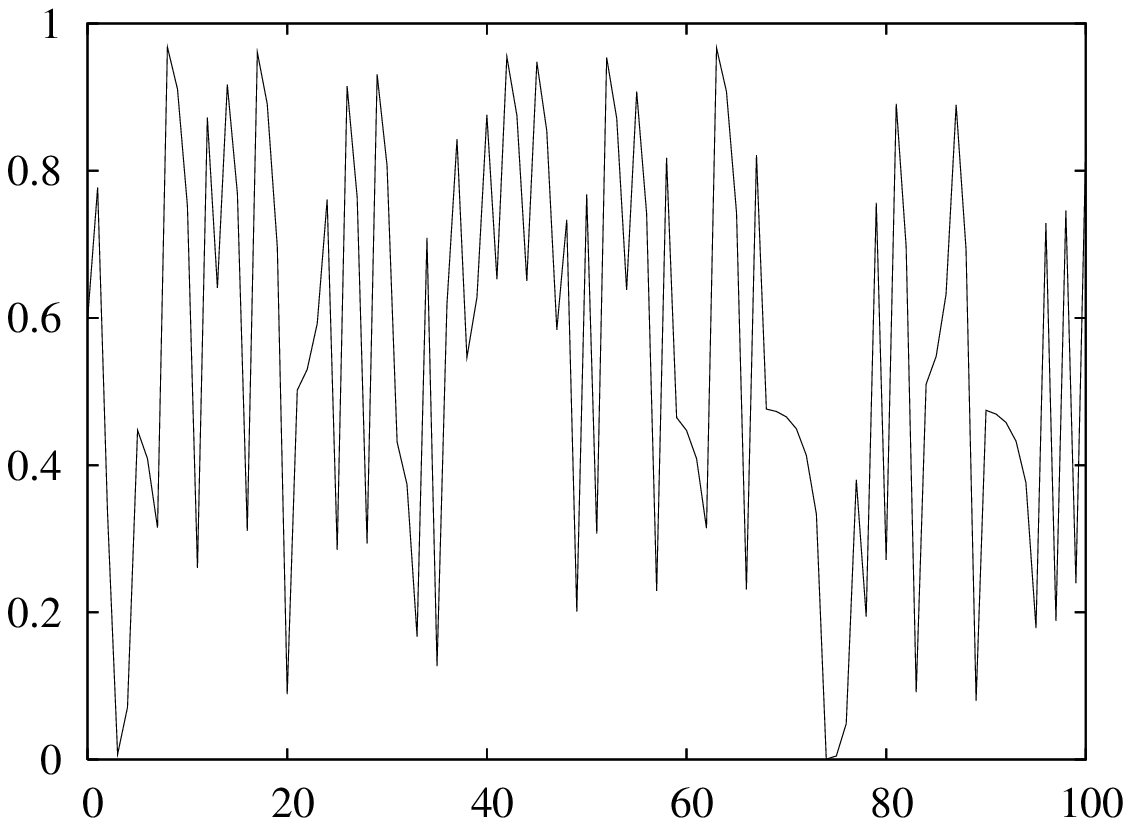}}&
   \resizebox{60mm}{40mm}{\includegraphics{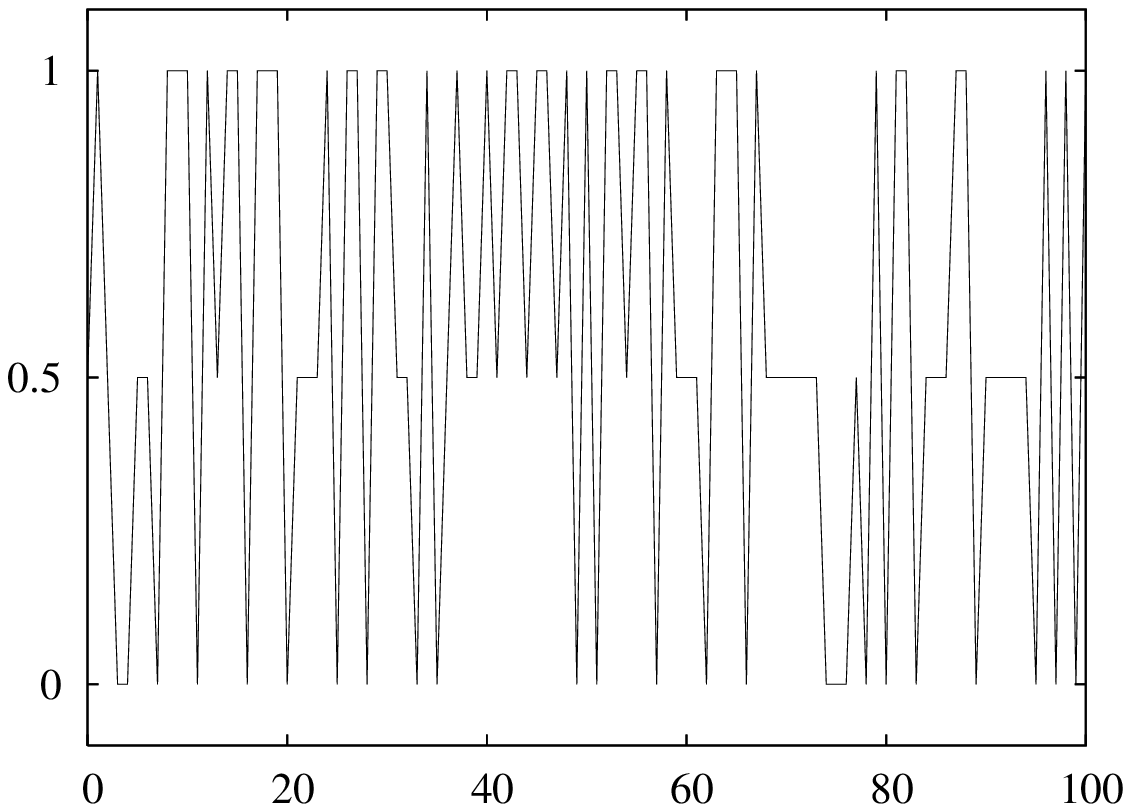}}\\
    \begin{picture}(0,0)(0,0)
      \put(0,310){(a)}
      \put(174,310){(b)}
      \put(-88,133){(c)}
      \put(5,18){{\large $n$}}
      \put(176,18){{\large $n$}}
      \put(88,224){\rotatebox{90}{\large $\rho(x)$}}
      \put(178,138){\large $x$}
      \put(5,138){\large $x_n$}
      \put(-88,70){\rotatebox{90}{\large $x_n$}}
      \put(90,70){\rotatebox{90}{\large $u_n$}}
      \put(-88,214){\rotatebox{90}{\large $x_{n+1}$}}
    \end{picture}
\end{tabular}
  \end{center}
  \caption{\label{fig:fig3}
Comparison of the  results by the Markov process and the Kalman map for $N=3$.  
(a) Kalman map corresponding to the Markov process with 
the transition matrix given in Eq.~(\ref{eq:H}).  
(b) Comparison of the exact invariant density (thick line)
with the approximate invariant density (thin line) in terms of unstable 15-periodic orbits by making use of the formula (\ref{eq:upo-density}).
(c) Chaotic time series generated by the mapping dynamics (Left) in (a) and 
the state evolution (Right) corresponding to the figure (a).  The stochastic 
evolution is generated by the equivalent Kalman dynamics.   
}
\end{figure*}

Hereafter we will compare the exact time correlation 
functions with those approximately determined by periodic orbits for $N=2$ and $3$.
The transition matrices $\hat{H}$'s for $N=2$ and $3$ under study are 
respectively
\begin{eqnarray}
\left(
\begin{array}{@{\,}cccc@{\,}}
0.41 & 0.83 \\
0.59 & 0.17 
\end{array}
\right) , 
\label{eq:H}
\left(
\begin{array}{@{\,}cccc@{\,}}
0.10 & 0.24 & 0.33 \\
0.33 & 0.45 & 0.31 \\
0.57 & 0.31 & 0.36
\end{array}
\right) . \\ \nonumber 
\end{eqnarray}
We use the values $a^{(1)}=0$ and $a^{(2)}=1$ for the $N=2$ case, 
$a^{(1)}=0, a^{(2)}=\frac{1}{2}$ and $a^{(3)}=1$ for the $N=3$ case.
The results are shown in Fig.~\ref{fig:fig2} and Tab.~\ref{tab:table1} for $N=2$ and Fig.~\ref{fig:fig3} and Tab.~\ref{tab:table2} for $N=3$.  
Figure \ref{fig:fig2} (a) and Fig. \ref{fig:fig3} (a) show the strange attractors of the one-dimensional maps corresponding to the transition matrices in Eq.~(\ref{eq:H}).  
Figure \ref{fig:fig2} (b) and Fig. \ref{fig:fig3} (b) are the comparison of the exact invariant densities (thick lines)
with the approximate ones (thin lines) in terms of unstable 15-periodic orbits by use of formula Eq. (\ref{eq:upo-density}).
Figure \ref{fig:fig2} (c) and Fig. \ref{fig:fig3} (c) are the chaotic time series (Left) and the switching between the states 
(Right) which is generated according to the chaotic time series.  
The comparison between the time correlation function obtained 
with the Markov method ($M=0$ and $1$ respectively for $N=2$ and $3$) and 
the unstable periodic orbits with the exact result 
from Eq.~(\ref{eq:exact-correlation}) is given in Tables \ref{tab:table1} and \ref{tab:table2}.  
One finds that the Markov method with unstable periodic orbits works quite well.  
The above results imply that 
{\it the statistical quantities in the stochastic 
process can be determined in terms of unstable periodic orbits 
embedded in the Kalman map}.
\begin{table}[ht]
  \begin{center}
    \begin{tabular}{c|c|c}
      \hline \hline
      n & exact  & approximate \\
      \hline
      0 & 0.2425 & 0.2335\\
      1 & -0.1027 & -0.0984\\
      2 & 0.0421 & 0.0434\\
      3 & -0.0193 & -0.0173\\
      4 & 0.0097 & 0.0078\\
      5 & -0.0044 & -0.0041\\
      6 & 0.0025 & 0.0019\\
      7 & 0.0003 & -0.0004\\
      8 & 0.0000 & 0.0000\\
      \hline
    \end{tabular}
  \end{center}
  \caption{\label{tab:table1}
 Comparison of the exact time correlation function $C_n$ from the formula 
(\ref{eq:exact-correlation}) with the approximate one by the Markov method with $M=0$ for $N=2$.
The long time average for the latter was replaced 
by the ensemble average with unstable periodic orbits.  See Eq.~(\ref{eq:upo-density}) and Fig.~{\ref{fig:fig2}}(b).}
\end{table}

\begin{table}[ht]
  \begin{center}
    \begin{tabular}{c|c|c}
      \hline \hline
      n & exact & approximate \\
      \hline
      0 & 0.1538 & 0.1538\\
      1 & -0.0327 & -0.0321\\
      2 & 0.0085 & 0.0093\\
      3 & -0.0028 & -0.0021\\
      4 & 0.0007 & -0.0004\\
      5 & -0.008 & -0.0003\\
      6 & 0.0006 & 0.0001\\
      7 & 0.0000 & 0.0000\\
      8 & 0.0000 & 0.0000\\
      \hline
    \end{tabular}
  \end{center}
  \caption{\label{tab:table2}
 Comparison of the exact time correlation function $C_n$ from the formula 
(\ref{eq:exact-correlation}) with approximate one by the Markov method with $M=1$ for $N=3$.
The long time average for the latter was replaced 
by the ensemble average with unstable periodic orbits.  See Eq.~(\ref{eq:upo-density}) and Fig.~{\ref{fig:fig3}}(b).}
\end{table}

\section{Concluding remarks and Discussions}
In the present paper, we showed that the large deviation 
deviation statistical quantities of a discrete time, finite state Markov 
process precisely coincides with that obtained by the Kalman 
map corresponding to the Markov process.  
The chaotic dynamics is self-generated and has an inner dynamics.  
In this sense, although the Kalman dynamics generates the 
stochastic process under consideration, two dynamics are different. 
Nevertheless, if one observes the dynamics of 
the coarse-grained variable, namely the variable $u(x)$ being 
independent of $x$ if $\beta_{(j-1)N}<x\le \beta_{jN}$ for each label 
$j$, provided one cannot distinguish the dynamics of 
the chaotic dynamics and the stochastic dynamics, then two dynamics 
give rigorously same results on statistical quantities, i.e., 
the invariant probability, double-time correlation functions and the 
large deviation statistical quantities.
\par
Differences between the stochastic process and the Kalman 
dynamics are caused by the fact that the chaotic 
variable $x_n$ is continuous in contrast to that the states of the 
present stochastic process are discrete.  Furthermore, 
the Lyapunov exponent is determined for the chaotic dynamics, while 
it cannot be determined for the stochastic process.  
The Lyapunov exponent $\lambda =\langle \ln |f'(x)| 
\rangle =\int \rho (x)\ln |f'(x)|dx$ for the Kalman 
map is easily calculated as
\begin{eqnarray}
\lambda =\sum_{j=1}^NP_*^{(j)}\sum_{k=1}^NH_{kj}\ln H_{kj}^{-1}.
\label{eq:lyapunov}
\end{eqnarray}
Although the Lyapunov exponent is the key concept of chaotic 
system and cannot be defined in a stochastic process in a 
conventional sense, the Lyapunov exponent of the Kalman 
dynamics is fully determined by the quantities 
contained in the stochastic process.  
In this sense, the quantity (\ref{eq:lyapunov}) can be called the 
Lyapunov exponent of the stochastic process (1).  
One can conclude that the origins of the unpredictability 
in the finite-state Markov stochastic process and the 
Kalman dynamics, more exactly speaking, the chaotic 
dynamics are same.  
It is worth while to note that Eq.~(\ref{eq:lyapunov}) is identical to 
the equality between the Lyapunov exponent and the 
Kolmogorov-Sinai entropy of one-dimensional map~\cite{O93,B65,CFS82}.  
\par
Since a Markov stochastic process can be generated by the 
Kalman map, statistical quantities of the stochastic process are 
determined by the chaotic dynamics.  
By making use of simple examples, 
we showed that {\it dynamical
quantities of the stochastic process can be well approximated in 
terms of unstable periodic orbits of the Kalman dynamics} in 
Sec.~III.  
\par
Recently, statistical quantities of the turbulence can be approximated with
an admissible unstable periodic orbit~\cite{KK01}.
Furthermore the time correlation function of chaotic 
dynamics can be well approximated with an appropriate unstable 
periodic orbit embedded in the attractor~\cite{KF06}.
In the case of Kalman dynamics, 
we showed in Appendix D that the approximation of time correlations with an appropriate unstable periodic orbit is well done.
It is found that the admissible unstable periodic orbit have a passing rate which is similar to the invariant density of the Kalman map.
\par
In closing the paper, let us discuss the applicability of 
Kalman dynamics to more general stochastic processes.  
As discussed in the present 
paper, the Kalman map precisely explains the statistics of 
finite state discrete time Markov process.  However, the Kalman map 
is a quite special type of mapping dynamics even the dynamics is 
restricted in mapping system in the sense that the Kalman map is 
piecewise linear and everywhere hyperbolic.  
In physical systems observed in experiments, 
mapping dynamics are usually neither piecewise nor everywhere 
hyperbolic.  This fact implies that physically observed stochastic 
processes generically cannot be described by finite-state Markov 
stochastic process.  It is quite interesting and important to 
study the possibility to construct a chaotic dynamics 
which describe more complicated stochastic dynamics such as 
continuous state and continuous time stochastic process.
\section*{Acknowledgments}
The authors are grateful to Hiroki Hata 
for valuable discussions.   
One of the authors (M.U.K.) thanks the Grant-in-Aid for JSPS Fellows.
This study was partially 
supported by the 21st Century COE program ``Center of Excellence 
for Research and Education on Complex Functional Mechanical 
Systems'' at Kyoto University.
\appendix
\section{Kalman Maps for two and three state processes}
In this appendix, examples of the Kalaman map are shown for $N=2$ and
3. 
One should note that the dynamics $x_{n+1}=f(x_n)$ with the mapping
function (12)
shows a chaotic behavior since the local expansion rate $\ln |f'(x)|$ of
the
mapping function is everywhere positive.  Therefore, the mapping
system (12) turns out to be hyperbolic.
\par
Let us first consider the two-state stochastic process.
The positions $\beta_{\underline{j}}\ (\underline{j}=1,2,3)$ 
are obtained as
\begin{eqnarray}
\beta_1=\frac{H_{11}}{2},\ \beta_2=\frac{1}{2},\ 
\beta_3=\frac{1}{2}(1+H_{12}).
\end{eqnarray}
\par
The mapping function of the Kalman map corresponding to
the above is given by
\begin{eqnarray}
f(x)= \begin{cases}
      H_{11}^{-1}x & \text{($0< x\le \beta_1$)}, \\
      \beta_2+H_{21}^{-1}(x-\beta_1) & \text{($\beta_1< x\le \beta_2$)},
\\
      H_{12}^{-1}(x-\beta_2) & \text{($\beta_2< x\le \beta_3$)}, \\
      \beta_2+H_{22}^{-1}(x-\beta_3 ) & \text{($\beta_3< x\le 1$)}.
      \end{cases}
\end{eqnarray}
The above function is drawn in Fig.~3.
\begin{figure}[h]
\begin{center}
  \includegraphics[width=.5\linewidth,angle=0]{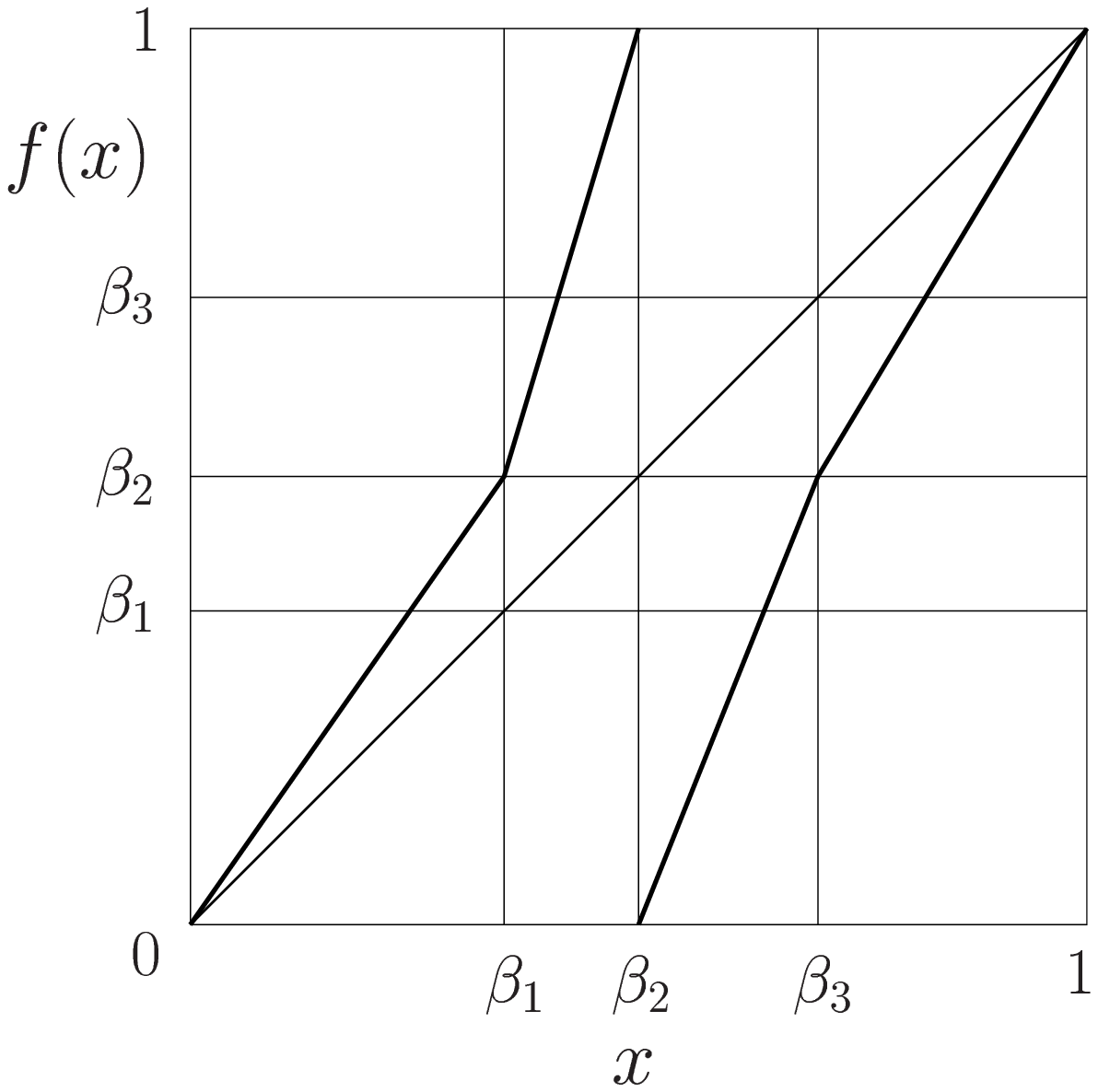}
  \caption{\label{fig5}
  Kalman map constructed from the
two-state stochastic process.  The positions $\beta_{\underline{j}}$'s
are
given in Eq.~(A1).
    }
\end{center}
\end{figure}
\par
For the three-state stochastic process,
the positions $\beta_{\underline{j}}\ (\underline{j}=1,2,\cdots ,8)$
are
obtained as
\begin{eqnarray}
\beta_1&=&\frac{H_{11}}{3},\ \beta_2=\frac{1}{3}(H_{11}+H_{21}),\
\beta_3=\frac{1}{3},\nonumber \\
\beta_4&=&\frac{1}{3}(1+H_{12}),\
\beta_5=\frac{1}{3}(1+H_{12}+H_{22}),\ \beta_6=\frac{2}{3},
\nonumber \\
\beta_7&=&\frac{1}{3}(2+H_{13}),\
\beta_8=\frac{1}{3}(2+H_{13}+H_{23}). \nonumber \\
\ \ \
\end{eqnarray}
The mapping function of the constructed one-dimensional map is given by
\begin{eqnarray}
f(x)= \begin{cases}
      H_{11}^{-1}x & \text{($0< x\le \beta_1$)}, \\
      \beta_3+H_{21}^{-1}(x-\beta_1) & \text{($\beta_1< x\le \beta_2$)},
\\
      \beta_6+H_{31}^{-1}(x-\beta_2) & \text{($\beta_2< x\le \beta_3$)},
\\
      H_{12}^{-1}(x-\beta_3)         & \text{($\beta_3< x\le \beta_4$)},
\\
      \beta_3+H_{22}^{-1}(x-\beta_4) & \text{($\beta_4< x\le \beta_5$)},
\\
      \beta_6+H_{32}^{-1}(x-\beta_5) & \text{($\beta_5< x\le \beta_6$)},
\\
      H_{13}^{-1}(x-\beta_6)         & \text{($\beta_6< x\le \beta_7$)},
\\
      \beta_3+H_{23}^{-1}(x-\beta_7) & \text{($\beta_7< x\le \beta_8$)},
\\
      \beta_6+H_{33}^{-1}(x-\beta_8) & \text{($\beta_8< x\le 1$)}.
      \end{cases}
\end{eqnarray}
The function $f(x)$ is drawn in Fig.~4.
\begin{figure}[h]
\begin{center}
  \includegraphics[width=.5\linewidth,angle=0]{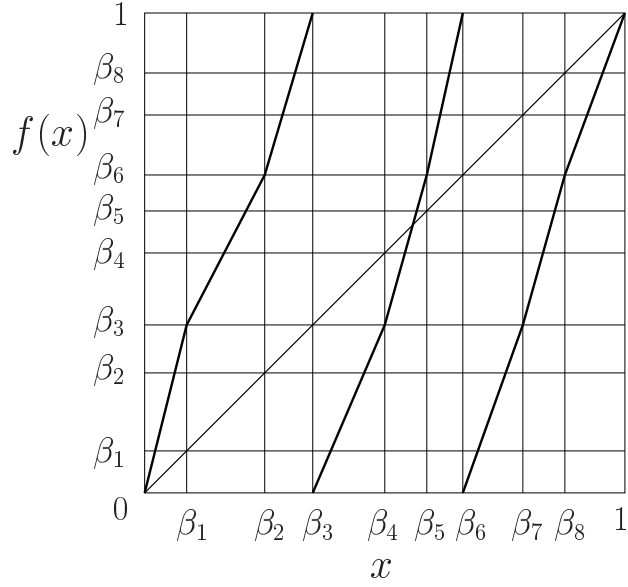}
  \caption{\label{fig6}
Kalman map constructed from the
three-state stochastic process.  The positions $\beta_{\underline{j}}$'s
are
given in Eq.~(A3). 
    }
\end{center}
\end{figure}
\section{Proofs of the statements in sec. II}
\subsection{Invariant probability density}
Let us consider the chaotic dynamics $x_{n+1}=f(x_n)$ with the
piecewise linear mapping
function (12).  The probability density $\rho_n(x)$ obeys the
evolution equation $\rho_{n+1}(x)={\cal H}\rho_n(x)$ with the
Frobenius-Perron operator ${\cal H}$. 
One can show that the invariant density $\rho (x)(={\cal H}\rho (x))$
is expanded as
\begin{eqnarray}
\rho (x)=\sum_{\underline{j}=1}^{N^2}\underline{c}_{\underline{j}}
\underline{I}_{\underline{j}}(x).
\end{eqnarray}
Here the function $\underline{I}_{\underline{j}}(x)$ is defined as
\begin{eqnarray}
\underline{I}_{\underline{j}}(x)=
\begin{cases}
1 &  \text{for $\beta_{\underline{j}-1}<x\le \beta_{\underline{j} }$},\\
0 &  \text{otherwise},
\end{cases}
\end{eqnarray}
$(\underline{j}=1,2,\cdots , N^2)$. 
The coefficients $\underline{c}_1,\underline{c}_2, \cdots
,\underline{c}_{N^2}$
are determined as follows.  First, note the relation
\begin{eqnarray}
{\cal H}\underline{I}_{\underline{j}}(x)=\sum_{\mu }
\frac{\underline{I}_{\underline{j}}(y_{\mu})}{|f'(y_{\mu })|}=
\sum_{\underline{k}=1}^{N^2}\underline{I}_{\underline{k}}(x)
\underline{H}_{\underline{k}\underline{j}},
\end{eqnarray}
where $\underline{H}_{\underline{k}\underline{j}}$ is the
$\underline{k}\underline{j}$ element of an $x$-independent $N^2\times
N^2$
matrix $\hat{\underline{H}}$.  The matrix $\hat{\underline{H}}$ is
obtained as follows. 
By putting
$\underline{j}=(j-1)N+j' \ , \ \underline{k}=(k-1)N+k'$ with
$
1\le j\le N,\ 1\le k\le N, \ 1\le j'\le N$ and $\ 1\le k'\le N 
$
and
by noting $\underline{I}_{\underline{j}}(y_{\mu})/|f'(y_{\mu })|
=H_{j'j}\underline{I}_{\underline{j}}(y_{\mu})$, Eq.~(B3) is written
as
\begin{eqnarray}
H_{j'j}\sum_{\mu }
\underline{I}_{\underline{j}}(y_{\mu })=
\sum_{\underline{k}=1}^{N^2}\underline{I}_{\underline{k}}(x)
\underline{H}_{\underline{k}\underline{j}}.
\end{eqnarray}
Multiplying $\underline{I}_{\underline{k}}(x)$ to Eq.~(B3) and
integrating it over $x$, we obtain
\begin{eqnarray}
\underline{H}_{\underline{k}\underline{j}}&=&N\frac{H_{j'j}}{H_{k'k}}
\int \underline{I}_{(k-1)N+k'}(x)
\sum_{\mu } \underline{I}_{(j-1)N+j'}(y_{\mu }(x))dx \nonumber \\
&=&H_{kj}\delta_{kj'},
\end{eqnarray}
where we gave used $\underline{I}_{(k-1)N+k'}(x)\sum_{\mu }
\underline{I}_{(j-1)N+j'}
(y_{\mu }(x))=\delta_{j'k}\underline{I}_{(k-1)N+k'}(x)$. 
\par
The explicit form of the $N^2 \times N^2$ matrix $\hat{\underline{H}}$ is given by
\begin{eqnarray}
\hat{\underline{H}}=\hspace{15cm} \nonumber \\
\left(
\begin{array}{@{\,}ccccc|ccccc|c|ccccc@{\,}}
H_{11} & 0 & \ldots   & 0 & 0 & H_{12} & 0 & \ldots   & 0 & 0 & \ldots &
H_{1N} & 0 & \ldots   & 0 & 0 \\
H_{11} & 0 & \ldots   & 0 & 0 & H_{12} & 0 & \ldots   & 0 & 0 & \ldots &
H_{1N} & 0 & \ldots   & 0 & 0 \\
\vdots & \vdots & \ddots & \vdots & \vdots & \vdots & \vdots & \ddots &
\vdots & \vdots & \vdots &\vdots &\vdots &\ddots & \vdots & \vdots  \\
H_{11} & 0 & \ldots   & 0 & 0 & H_{12} & 0 & \ldots   & 0 & 0 & \ldots &
H_{1N} & 0 & \ldots   & 0 & 0 \\ \hline
0 & H_{21} & \ldots   & 0 & 0 & 0 & H_{22} & \ldots   & 0 & 0 & \ldots &
0 & H_{2N} & \ldots   & 0 & 0 \\
0 & H_{21} & \ldots   & 0 & 0 & 0 & H_{22} & \ldots   & 0 & 0 & \ldots &
0 & H_{2N} & \ldots   & 0 & 0 \\
\vdots & \vdots & \ddots & \vdots & \vdots & \vdots & \vdots & \ddots &
\vdots & \vdots & \vdots &\vdots &\vdots &\ddots & \vdots & \vdots  \\
0 & H_{21} & \ldots   & 0 & 0 & 0 & H_{22} & \ldots   & 0 & 0 & \ldots &
0 & H_{2N} & \ldots   & 0 & 0 \\ \hline
\ \  & \vdots & \ \ & \ \ & \ \ &  \ \  & \vdots & \ \ & \ \ & \ \ & 
\ddots &\ \  & \vdots & \ \ & \ \ & \ \   \\ \hline
0 & 0 & \ldots   & 0 & H_{N1}& 0 & 0 & \ldots &0 &H_{N2} & \ldots & 0  &
0 & \ldots & 0 & H_{NN} \\
0 & 0 & \ldots   & 0 & H_{N1}& 0 & 0 & \ldots &0 &H_{N2} & \ldots & 0  &
0 & \ldots & 0 & H_{NN} \\
\vdots & \vdots & \ddots & \vdots & \vdots & \vdots & \vdots & \ddots &
\vdots & \vdots & \vdots &\vdots &\vdots &\ddots & \vdots & \vdots  \\
0 & 0 & \ldots   & 0 & H_{N1}& 0 & 0 & \ldots &0 &H_{N2} & \ldots & 0  &
0 & \ldots & 0 & H_{NN}                  
\end{array}
\right) . \nonumber \\
\ \
\end{eqnarray}
\par
With the expression (B1) and the relation (B3), we get
\begin{eqnarray}
{\cal H}\rho
(x)=\sum_{\underline{j}=1}^{N^2}\underline{c}_{\underline{j}}
\sum_{\underline{k}=1}^{N^2}
\underline{I}_{\underline{k}}(x)
\underline{H}_{\underline{k} \underline{j}}.
\end{eqnarray}
Equating Eqs.~(B1) and (B7), one obtains
\begin{eqnarray}
\underline{c}_{\underline{j}}=\sum_{\underline{k}=1}^{N^2}
\underline{H}_{\underline{j}\underline{k}}\underline{c}_{\underline{k}}.
\end{eqnarray}
By noting the relation (B5), Eq.~(B8) is rewritten as
\begin{eqnarray}
\underline{c}_{(j-1)N+j'}=\sum_{k=1}^NH_{jk}\underline{c}_{(k-1)N+j},
\end{eqnarray}
where we put
$\underline{j}=(j-1)N+j' \ , \ \underline{k}=(k-1)N+k',
1\le j\le N,\ 1\le k\le N$ and $\ 1\le j'\le N, \ 1\le k'\le N$. 
Since the r.h.s. of Eq.~(B9) is independent of $j'$,
we find that $\underline{c}_{(j-1)N+j'}$ is free from $j'$. 
Therefore, by putting
$
\underline{c}_{(j-1)N+1}=\underline{c}_{(j-1)N+2}=
\underline{c}_{(j-1)N+3}=\cdots =\underline{c}_{(j-1)N+N}\equiv c_j
$,
Eq.~(B9) is written as
\begin{eqnarray}
c_j=\sum_{k=1}^NH_{jk}c_k,
\end{eqnarray}
and the probability density $\rho (x)$ given in Eq.~(B1) is rewritten as

\begin{eqnarray}
\rho (x)=\sum_{j=1}^Nc_jI_j(x),
\end{eqnarray}
where we have defined
\begin{eqnarray}
I_j(x)=
\begin{cases}
1 & \text{for $\beta_{(j-1)N}<x\le \beta_{jN}$}, \\
0 & \text{otherwise}.
\end{cases}
\end{eqnarray}
In order that Eq.~(B10) agrees with the result (4) in the
stochastic process, we need $c_j=\alpha P_*^{(j)}$, where $\alpha $ is
independent of $j$.  Furthermore, noting the normalization condition of
the
probability density $\rho (x)$, we find
$c_j=NP_*^{(j)}$, $(j=1,2,\cdots ,N)$. 
Equation (B10) thus turns out to be identical
with Eq.~(4) in the stochastic process.
\subsection{Double-time correlation function}
The equivalence of the double time correlation functions derived from
the
Markov process (1) and that from the corresponding Kalman map
was recently shown by Kohda and Fujisaki~\cite{KF99}.  The following
discussion
is the simplified one of their proof. 
The time correlation function $C_n^{\rm K}$ for a function $u(x_n)$,
where $u(x)$
takes the value $a^{(j)}$ if $\beta_{(j-1)N}<x\le \beta_{jN}$, is
rewritten as
\begin{eqnarray}
C_n^{\rm K}=\int \delta u(x_n)\delta u(x)\rho (x)dx.
\end{eqnarray}
By defining the quantities
$\underline{a}^{(\underline{k})}$'s by
$\underline{a}^{(\underline{k})}=a^{(k)}$ for
$\underline{k}=(k-1)N+k',
(1\le k\le N, 1\le k'\le N)$ and $\delta
\underline{a}^{(\underline{k})}=
\underline{a}^{(\underline{k})}-\langle u\rangle $ and inserting
the expression (B1) and
$\delta u(x)=\sum_{\underline{k}=1}^{N^2}\delta
\underline{a}^{(\underline{k})}
\underline{I}_{\underline{k}}(x)$ into Eq.~(B13), one gets
\begin{eqnarray}
C_n^{\rm K}&=&\int  \rho (x)\delta u(x){\cal L}^n\delta u(x) dx 
\nonumber \\
  &=& \int  \delta u(x){\cal H}^n[\rho (x)\delta u(x)] dx  \nonumber
\\
  &=&
\sum_{\underline{j}=1}^{N^2}\sum_{\underline{k}=1}^{N^2}
\underline{c}_{\underline{j}}\delta \underline{a}^{(\underline{j})}
\delta \underline{a}^{(\underline{k})}\int
\underline{I}_{\underline{k}}(x){\cal H}^n
\underline{I}_{\underline{j}}(x)dx,
\end{eqnarray}
where ${\cal L}$ is the time evolution operator defined by
${\cal L}G(x)=G(f(x))$.  By noting
\begin{eqnarray}
{\cal
H}^n\underline{I}_{\underline{j}}(x)=\sum_{\underline{k}=1}^{N^2}
\underline{I}_{\underline{k}}(x)(\hat{\underline{H}}^n)_{\underline{k}
\underline{j}},
\end{eqnarray}
the time correlation function is expanded as
\begin{eqnarray}
C_n^{\rm
K}=\sum_{\underline{j}=1}^{N^2}\sum_{\underline{k}=1}^{N^2}\Delta_{\underline{k}}
\delta
\underline{a}^{(\underline{k})}(\hat{\underline{H}}^n)_{\underline{k}
\underline{j}}
\delta \underline{a}^{(\underline{j})}\underline{c}_{\underline{j}},
\end{eqnarray}
where $\Delta_{\underline{k}}\equiv \int
\underline{I}_{\underline{k}}(x)dx$. 
Putting $\underline{j}=(j-1)N+j'$ and $\underline{k}=(k-1)N+k'$ with
$1\le j\le N, 1\le k\le N,1\le j'\le N$ and $1\le k'\le N$, we get
$\Delta_{\underline{k}}=\beta_{(k-1)N+k'}-\beta_{(k-1)N+k'-1}=H_{k'k}/N$, and
therefore $\sum_{k'=1}^{N}\Delta_{(k-1)N+k'}=\beta_{kN}-\beta_{(k-1)N}=N^{-1}$. 
These relations lead to
\begin{eqnarray}
(\hat{\underline{H}}^n)_{\underline{k}\underline{j}}
=(\hat{H}^{n-1})_{kj'}H_{j'j},
\end{eqnarray}
$(n=1,2,3,\cdots )$.  Furthermore, since $\delta
\underline{a}^{(\underline{j})}
=\delta a^{(j)}$ and $\underline{c}_{\underline{j}}=NP_*^{(j)}$,
Eq.~(B16) is reduced to the time correlation function (5)
obtained in the stochastic process.
\subsection{Large deviation theoretical characteristic function}
The large deviation theoretical characteristic function for the time
series
$\{ u(x_m)\} $, where $u(x)$ takes the value $a^{(j)}$ if
$\beta_{(j-1)N}<x\le \beta_{jN}, (j=1,2,\cdots ,N)$.  The
characteristic
function is rewritten as
\begin{eqnarray}
Z^{\rm K}_q(n)=\int {\cal H}_q^n\rho (x)dx,
\end{eqnarray}
where ${\cal H}_q$ is the generalized (order-$q$) Frobenius-Perron
operator~\cite{ST86,FI87b} defined as
\begin{eqnarray}
{\cal H}_qG(x)\equiv {\cal H}\left[ e^{qu(x)}G(x) \right]
=\sum_{\mu }\frac{e^{qu(y_{\mu })}G(y_{\mu })}{|f'(y_{\mu })|}.
\end{eqnarray}
\par
Noting the relation
\begin{eqnarray}
{\cal H}_q\underline{I}_{\underline{j}}(x)=
\sum_{\underline{k}=1}^{N^2}\underline{I}_{\underline{k}}(x)
(\hat{\underline{H}}_q)_{\underline{k}\underline{j}}
\end{eqnarray}
with the matrix $\hat{\underline{H}}_q$ whose
$\underline{k}\underline{j}$ element is defined by
\begin{eqnarray}
(\hat{\underline{H}}_q)_{\underline{k}\underline{j}}=
\underline{H}_{\underline{k}\underline{j}}\exp
[q\underline{a}^{(\underline{j})}],
\end{eqnarray}
$(\hat{\underline{H}}_{q=0}=\hat{\underline{H}})$, one obtains
\begin{eqnarray}
{\cal H}_q^n\underline{I}_{\underline{j}}(x)=
\sum_{\underline{k}}\underline{I}_{\underline{k}}(x)
(\hat{\underline{H}}_q^n)_{\underline{k}\underline{j}}.
\end{eqnarray}
Therefore, since
$
{\cal H}_q^n\rho (x)=\sum_{\underline{j}}\sum_{\underline{k}}
\underline{c}_{\underline{j}}\underline{I}_{\underline{k}}(x)
(\hat{\underline{H}}_q^n)_{\underline{k}\underline{j}}
$,
we obtain
$
Z_q^{\rm K}(n)=\sum_{\underline{j}}\sum_{\underline{k}}
\underline{c}_{\underline{j}}\Delta_{\underline{k}}
(\hat{\underline{H}}_q^n)_{\underline{k}\underline{j}}
$ 
If we put $\underline{j}=(j-1)N+j',\underline{k}=(k-1)N+k'$ with
$1\le j\le N, 1\le j'\le N, 1\le k\le N$ and $1\le k'\le N$, then
noting $\underline{c}_{\underline{j}}=c_j, \Delta_{\underline{k}}=
H_{k'k}/N$ and
\begin{eqnarray}
(\hat{\underline{H}}_q)_{\underline{k}\underline{j}}&=&
(\hat{H}_q)_{kj}\delta_{kj'}, \\
(\hat{\underline{H}}_q^n)_{\underline{k}\underline{j}}&=&
(\hat{H}_q^{n-1})_{kj'}
(\hat{H}_q)_{j'j},
\end{eqnarray}
where $( \hat{H}_q)_{kj}=H_{kj}e^{qa^{(j)}}$,
we obtain
$
Z_q^{\rm K}(n)=\sum_j\sum_k(\hat{H}_q^n)_{kj}\frac{c_j}{N}
$. 
By making use of $c_j=NP_*^{(j)}$, the above expression
coincides with Eq.~(9).
\par
As proved above, those of the present chaotic dynamics constructed in
the
preceding section gives the results precisely same as those of the
stochastic
process.  Therefore, the one-dimensional chaotic dynamics with the
mapping function (12) precisely simulates the stochastic process (1).
\section{Markov method for time correlation functions of one-dimensional
map}
We consider a chaotic one-dimensional map,
\begin{eqnarray}
x_{n+1} = f(x_n),
\end{eqnarray}
($n=0,1,2,\cdots )$.
The time series $\{ u_n\} $ under consideration is given by
$u_n=h\{ x_n\} $, where
$h\{ x\} $ is a unique scalar function of
$x$.  In terms of the time evolution
operator $\mathcal{L}$ defined by $\mathcal{L}G(x)=G(f(x))$, $u_n$
obeys the equation of motion $u_{n+1}=\mathcal{L}u_n$. 
The time correlation function $C_n=\langle u_nu_0
\rangle $, where $\langle \cdots \rangle$ denotes the long time average,
with $u_n$ being chosen such that $\langle u_n \rangle =0$,
can be obtained by the Markov method proposed in Ref.~\cite{F05} as follows.
\par
First, we introduce the vector variable
\begin{eqnarray}
{\mbox{\boldmath{$u$}}} \{ x\}
\equiv (h_1\{ x \} ,h_2
\{ x \} , \cdots , h_{M+1}
\{ x \})^T ,
\end{eqnarray}
where
$h_1\{ x \}$ is identical to $h\{ x \}$ under consideration.
$M$ is the number of new scalar variables, 
$h_2,h_3,\cdots , h_{M+1}$, and is assumed to be
suitably chosen.  The functions
$h_1,h_2, \cdots ,h_{M+1}$
are chosen so as to have vanishing means and have
components linearly independent of each other.
The vector variable ${\mbox{\boldmath{$u$}}}_n$ defined by
\begin{eqnarray}
{\mbox{\boldmath{$u$}}}_n
=\mathcal{L}^n{\mbox{\boldmath{$u$}}}_0,\ \ \
{\mbox{\boldmath{$u$}}}_0=
{\mbox{\boldmath{$u$}}}\{ x \},
\end{eqnarray}
obeys the equation of motion
${\mbox{\boldmath{$u$}}}_{n+1}=\mathcal{L}{\mbox{\boldmath{$u$}}}_n$.
\par 
With the projection operator method~\cite{FY78}, the above equation
can be written in the form of the Mori equation of
motion with a memory term.  If $M$ is
appropriately chosen, the contribution from the
memory term is expected to be small and can be ignored~\cite{F05}. 
With this approximation, the Mori equation reduces to
\begin{eqnarray}
{\mbox{\boldmath{$u$}}}_{n+1} \approx \hat{\zeta } 
{\mbox{\boldmath{$u$}}}_n +{\mbox{\boldmath{$g$}}}_n
\end{eqnarray}
with 
\begin{eqnarray}
\hat{\zeta }\equiv \langle [ \mathcal{L}
{\mbox{\boldmath{$u$}}}_0] {\mbox{\boldmath{$u$}}}^T_0 \rangle
\langle {\mbox{\boldmath{$u$}}}_0 {\mbox{\boldmath{$u$}}}^T_0 \rangle
^{-1}.
\end{eqnarray}
The fluctuating force ${\mbox{\boldmath{$g$}}}_n$ is orthogonal to
${\mbox{\boldmath{$u$}}}_0$, i.e.,
$
\langle  {\mbox{\boldmath{$g$}}}_n {\mbox{\boldmath{$u$}}}_0^T \rangle
=\hat{0}, \quad (n\ge 0)
$. 
By noting this property, the time correlation
matrix $\hat{C}_{n} \equiv \langle
{\mbox{\boldmath{$u$}}}_n{\mbox{\boldmath{$u$}}}^T_0 \rangle $
obeys $\hat{C}_{n+1} \approx \hat{\zeta } \hat{C}_n$, which
yields
\begin{eqnarray}
\hat{C}_n \approx \hat{\zeta }^n\hat{C}_0.
\end{eqnarray}
By noting $\hat{\zeta }=\hat{C}_1\hat{C}_0^{-1}$, 
the time correlation function $C_n$ is thus given by the
1-1 component of $\hat{C}_n$. 
The above approach to the time correlation function is called the
{\it Markov method}~\cite{F05,KF06}.
\section{Determination of time-correlation functions in terms of
one unstable periodic orbit}
In Sec. III, we determined time correlation functions of the Markov process in terms of many unstable periodic orbits embedded in the corresponding Kalman map.
In particular we describe time correlation functions with static quantities ($\hat{C}_0$ and $\hat{C}_1$) and those static quantities with many unstable periodic orbits.
However, for calculation of $\hat{C}_0$ and $\hat{C}_1$ we do not have to 
determine the invariant density in terms of so many unstable periodic orbits
because $\hat{C}_0$ and $\hat{C}_1$ include only low order momentums.
Here, as the simplest case, we determine dynamical correlations in terms of only single periodic orbit with a passing rate which is similar to invariant density.
Thus we use the approximation to determine $\hat{C}_0$ and $\hat{C}_1$ in terms of 
an appropriate unstable periodic orbit instead of the long-time average as
\begin{eqnarray}
\langle G(x)\rangle \approx \frac{1}{N_p}
\sum_{j=0}^{n-1}G(x_j^{(p)}),
\label{eq:single}
\end{eqnarray}
where $x_{j+N_p}^{(p)}=x_{j}^{(p)}$, $(j=1,2, \cdots ,N_p)$
being a period-$N_p$ unstable periodic orbit appropriately
chosen.
If this approximation holds,
then the dynamical correlation functions can be
approximately expanded
in terms of an unstable periodic orbit~\cite{KF06}.
\par
Hereafter we will compare the time correlation functions with those by
periodic orbit for N=2.
The transition matrices $\hat{H}$'s for $N=2$ under study is the same as Sec. III.
\begin{figure*}[th]
\begin{center}
\begin{tabular}{cc}
    \resizebox{100mm}{60mm}{\includegraphics{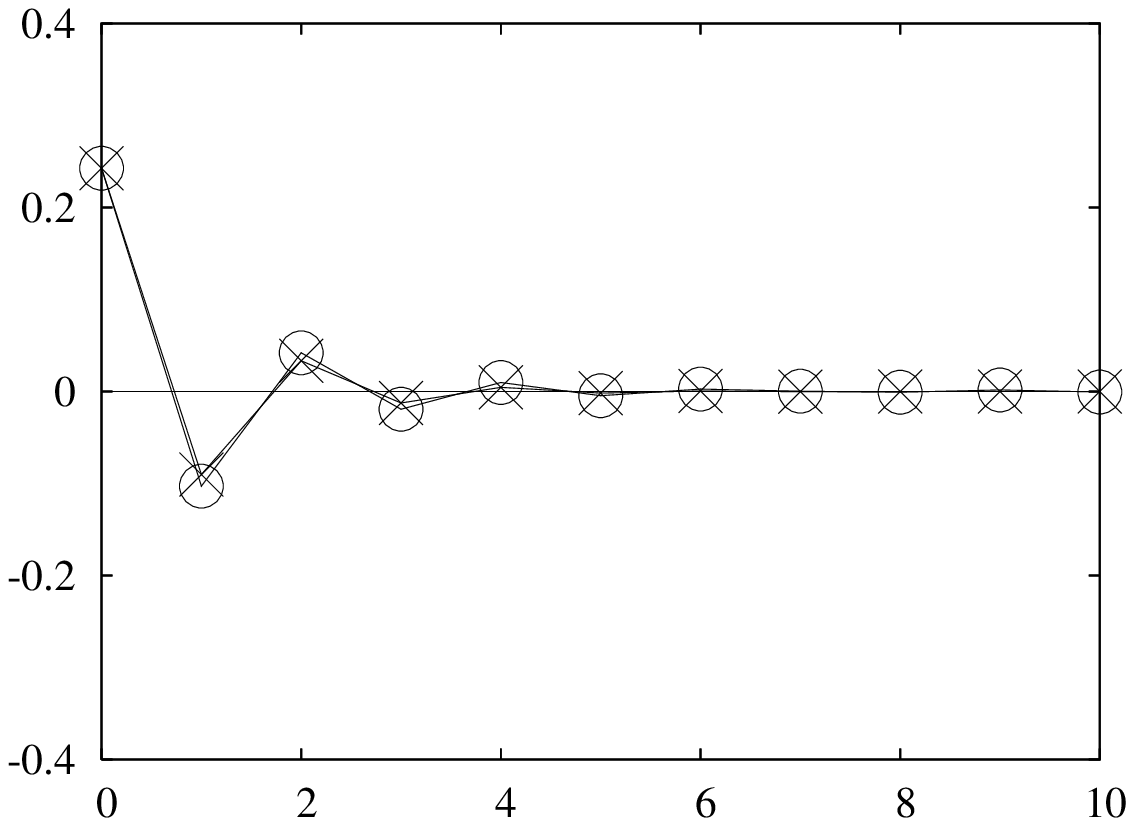}}&
    \resizebox{60mm}{60mm}{\includegraphics{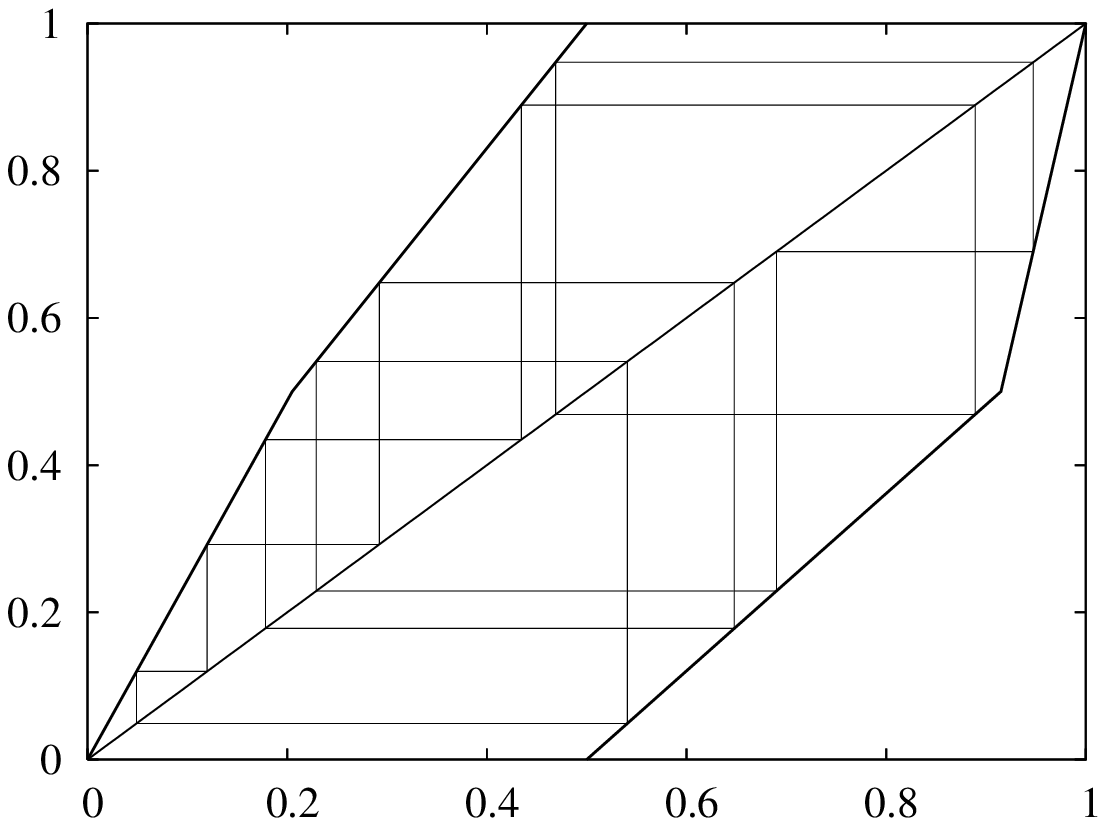}} \\
    \begin{picture}(0,0)(0,0)
      \put(4,188){(a)}
      \put(235,188){(b)}
      \put(-140,105){\rotatebox{90}{\Large ${C_n}$}}
      \put(6,12){{\large $n$}}
      \put(145,95){\rotatebox{90}{\large $x_{n+1}$}}
      \put(235,12){{\large $x_n$}}
    \end{picture}
\end{tabular}
  \end{center}
  \caption{\label{fig:upo12}
(a) Comparison of the exact time correlation function from the formula 
(15) ($\circ $) with the approximate one by the Markov method with $M=0$ 
($\times $).
The long time average for the latter was replaced 
by the average over the unstable periodic orbit shown 
in (b).  See Eq.~(\ref{eq:single}).
(b) The unstable periodic orbit with the period 12.
  }
\end{figure*}
The results are shown in Fig. (\ref{fig:upo12}).
Fig. (\ref{fig:upo12}) (a) shows the comparison between the time correlation functions obtained with the Markov method ($M=0$) and the unstable periodic orbit shown in (b) with the exact one, Eq. (\ref{eq:exact-correlation}).
One finds that even if only one unstable periodic orbit is used, 
one finds that the approximation is well done.

\end{document}